\documentclass[12pt]{article}

\usepackage{latexsym}
\usepackage{amsmath}
\usepackage{amssymb}
\catcode`\@=11
\newcount\hour
\newcount\minute
\newtoks\amorpm \hour=\time\divide\hour by 60\minute
=\time{\multiply\hour by 60 \global\advance\minute by-\hour}
\edef\standardtime{{\ifnum\hour<12 \global\amorpm={am}%
        \else\global\amorpm={pm}\advance\hour by-12 \fi
        \ifnum\hour=0 \hour=12 \fi
        \number\hour:\ifnum\minute<10
        0\fi\number\minute\the\amorpm}}
\edef\militarytime{\number\hour:\ifnum\minute<10
0\fi\number\minute}
\def\draftlabel#1{{\@bsphack\if@filesw {\let\thepage\relax
   \xdef\@gtempa{\write\@auxout{\string
      \newlabel{#1}{{\@currentlabel}{\thepage}}}}}\@gtempa
   \if@nobreak \ifvmode\nobreak\fi\fi\fi\@esphack}
        \gdef\@eqnlabel{#1}}
\def\@eqnlabel{}
\def\@vacuum{}
\def\marginnote#1{}
\def\draftmarginnote#1{\marginpar{\raggedright\scriptsize\tt#1}}
\def\draft{
        \pagestyle{plain}
        \overfullrule=2pt
        \oddsidemargin -.1truein
        \def\@oddhead{\sl \phantom{\today\quad\militarytime} \hfil
        \smash{\Large\sl DRAFT} \hfil \today\quad\militarytime}
        \let\@evenhead\@oddhead
        \let\label=\draftlabel
        \let\marginnote=\draftmarginnote
        \def\ps@empty{\let\@mkboth\@gobbletwo
        \def\@oddfoot{\hfil \smash{\Large\sl DRAFT} \hfil}
        \let\@evenfoot\@oddhead}
        \def\@eqnnum{(\theequation)\rlap{\kern\marginparsep\tt\@eqnlabel}%
        \global\let\@eqnlabel\@vacuum}  }

\renewcommand{\theequation}{\thesection.\arabic{equation}}
\renewcommand{\thefootnote}{\fnsymbol{footnote}}
\newcommand{\newsection}{    
\setcounter{equation}{0}\section}
\def\appendix#1{\addtocounter{section}{1}\setcounter{equation}{0}
\renewcommand{\thesection}{\Alph{section}}
\section*{Appendix \thesection\protect\indent \parbox[t]{11.15cm}{#1}}
\addcontentsline{toc}{section}{Appendix \thesection\ \ \ #1}}

\def \bi{\bibitem}
\def \la {\label}

\def\be{\begin{equation}}
\def\ee{\end{equation}}

\def\nat {{\natural}}

\catcode`\@=11

\def\bea{\begin{eqnarray}}
\def\eea{\end{eqnarray}}
\def\beann{\begin{eqnarray*}}
\def\eeann{\end{eqnarray*}}
\def\beq{\begin{equation}}
\def\eeq{\end{equation}}
\def\ba{\begin{array}}
\def\ea{\end{array}}
\def\ben{\begin{enumerate}}
\def\een{\end{enumerate}}

 \def \la {\label}
 \def\be{\begin{equation}}
\def\ee{\end{equation}}

\def \la {\label}

\font\mybb=msbm10 at 11pt

\def\bb#1{\hbox{\mybb#1}}

\def\bZ {\bb{Z}}
\def\bR {\bb{R}}

\def\bC {\bb{C}}

\def\e  {\epsilon}

\def \ee {\epsilon}

\def \bi{\bibitem}

\def \G {\Gamma}

\def\be{\begin{equation}}
\def\ee{\end{equation}}

\def \bi {\bibitem}
\def \la{\label}

\begin{document}
\date{November 2002}
\begin{titlepage}
\begin{center}
\vspace*{3.50cm}
{\Large \bf Invariant Killing spinors in 11D  and type II supergravities}\\[.2cm]

\vspace{1.5cm}
 {\large  U.~Gran$^1$, J.~Gutowski$^2$ and   G.~Papadopoulos$^3$  }

\vspace{0.5cm}

${}^1$ Fundamental Physics\\
Chalmers University of Technology\\
SE-412 96 G\"oteborg, Sweden\\

\vspace{0.5cm}
${}^2$ DAMTP, Centre for Mathematical Sciences\\
University of Cambridge\\
Wilberforce Road, Cambridge, CB3 0WA, UK

\vspace{0.5cm}
${}^3$ Department of Mathematics\\
King's College London\\
Strand\\
London WC2R 2LS, UK\\

\end{center}

\vskip 1.5 cm
\begin{abstract}
We present all isotropy groups and associated $\Sigma$ groups, up to discrete identifications of the component connected to the identity,
of spinors of eleven-dimensional and type II supergravities. The $\Sigma$ groups are products
of a $Spin$ group and an $R$-symmetry group of a suitable lower dimensional supergravity theory. Using the case
 of $SU(4)$-invariant spinors as a paradigm,   we demonstrate that the $\Sigma$ groups, and so the $R$-symmetry groups of lower-dimensional
 supergravity theories arising from compactifications, have disconnected components. These lead to discrete
 symmetry groups reminiscent of $R$-parity.
We examine the role of disconnected
components of the $\Sigma$ groups in the choice of Killing spinor representatives and in the context of compactifications.

\end{abstract}
\end{titlepage}
\newpage
\setcounter{page}{1}
\renewcommand{\thefootnote}{\arabic{footnote}}
\setcounter{footnote}{0}

\setcounter{section}{0}
\setcounter{subsection}{0}
\newsection{Introduction}

Supersymmetric supergravity backgrounds can be categorized into two classes. One class are those backgrounds for which the
Killing spinors are invariant under some proper Lie subgroup of the gauge group of the associated theory, and another class are those
backgrounds for which the isotropy group of the Killing spinors is the identity.
Most of the known supergravity backgrounds belong to the former class, like the vacua of string compactifications
with or without fluxes, for a recent review see \cite{grana}, and those backgrounds that have applications in gauge theory/duality correspondences,
see e.g.~\cite{klebanov, volkov, nunez, atgp}. A notable exception
are the maximally and the near maximally supersymmetric backgrounds of ten- and eleven-dimensional supergravities which belong to
the latter class and have been classified in \cite{jfofgp} and \cite{iibm31, bandos, josepreons}, respectively.

The main aim of this paper is to initiate the classification of {\it all} supersymmetric eleven-dimensional and type II backgrounds for which the
Killing spinors are invariant under some proper  Lie subgroup $H$ of the gauge group. Some partial results
are already known. In eleven-dimensional supergravity, these include the $N=1$ backgrounds with $SU(5)$- and $Spin(7)\ltimes\bR^9$-invariant Killing spinors \cite{janpakis};
the $N=2$ backgrounds with $SU(5)$- and some $N=2$ and $N=4$ backgrounds with $SU(4)$-invariant Killing spinors \cite{ggp, syst};
$N=2$ backgrounds with $Spin(7)$- and $G_2\ltimes \bR^9$-, and $N=4$ backgrounds with $G_2$- and $SU(4)$-invariant Killing spinors, and
special cases of backgrounds with stability subgroups embeddable in $Spin(7)\ltimes \bR^9$  \cite{mac1}.
In IIB supergravity, the Killing spinor equations have been solved for $N=1$ backgrounds in \cite{iib1} and all
backgrounds with maximal number of $H$-invariant spinors have been classified in \cite{iib2}.
The method that we shall
use to solve this problem is based on spinorial geometry \cite{ggp} facilitated by the application
of the $\Sigma({\cal P})$ groups defined in \cite{hetb}.
The $\Sigma({\cal P})$ groups  are the subgroups of the gauge group of the supergravity theories
that leave the plane ${\cal P}$ spanned  by some spinors
invariant. The importance of the $\Sigma({\cal P})$ groups has been demonstrated  in  the classification of all supersymmetric backgrounds
of heterotic supergravity \cite{hetb}. In particular, the $\Sigma({\cal P})$ groups have been
used to find the solutions of the
gaugino and dilatino Killing spinor equations given a solution of the gravitino Killing spinor equation.

To find the geometry of all eleven-dimensional backgrounds that admit $H$-invariant Killing spinors,
we shall first identify all the subgroups $H\subset Spin(10,1)$ that leave some spinors invariant. These subgroups
will be given up to {\it discrete identifications} of the connected to the identity component and so they will be derived from a Lie algebra computation.
Partial lists of such groups have appeared elsewhere \cite{breakwave, ggp, null}. Here we shall give the complete set
relevant for eleven-dimensional supergravity which is tabulated in table 1.
In addition, in tables 2 and 3, we give explicitly the representatives of all the $H$-invariant
spinors for all isotropy groups.

Given  ${\cal P}_H$, the plane spanned by {\it all} $H$-invariant spinors of eleven-dimensional supergravity, $N_H={\rm dim}\,{\cal P}_H$, we compute
the $\Sigma({\cal P}_H)$ group for every $H$. These are again  determined up to ${\it discrete}$ identifications of the connected to the identity
component and the computation is Lie algebraic.
 The list of all $\Sigma({\cal P}_H)$ groups is presented in table 5.
 We find that all the $\Sigma({\cal P}_H)$ groups are products $Spin\times R$,
 where $Spin$ and $R$ can be identified with the $Spin$ and $R$-symmetry groups
 of a lower-dimensional supergravity theory.

 Using the case of eleven-dimensional backgrounds with $SU(4)$-invariant spinors  as a paradigm, we demonstrate that
 the $\Sigma({\cal P}_H)$ groups have {\it disconnected} components which are subgroups of the
 connected component $Spin^0(10,1)$ of $Spin(10,1)$. In a compactification senario on a manifold with an $H$-structure
 and with or without fluxes,
 this implies that
 the R-symmetry group of the associated lower-dimensional supergravity theory is disconnected. In the $SU(4)$ case, representatives
 of the disconnected components act as a {\it discrete symmetry}  via reflections on some components of the  frame of the compact internal
 space. This action leaves the
 metric invariant but changes the fluxes, the fermions  and  the (almost) complex structure $I$ of the internal space to $-I$.
 Such discrete transformations are reminiscent\footnote{The discrete symmetries have also a passing similarity with the Weyl subgroup
 \cite{pope}
 of the U-duality group though they cannot be directly identified because the former appears in $SU(4)$-structure
 compactifications while the latter appears in toric ones.} of $R$-parity transformations imposed
 to suppress the rate of decay of the proton in supersymmetric theories, for a review see \cite{rparity}.
   Since these discrete symmetries are remnants of the {\it restricted} Lorentz transformations of an
  eleven-dimensional theory, it is natural to argue that they must be symmetries of the lower-dimensional
 effective supergravity theories imposing restrictions on the couplings.

Furthermore, we  explain how
 the $\Sigma({\cal P}_H)$ can be used in eleven-dimensional supergravity to find the normal forms of $H$-invariant Killing spinors for backgrounds
 with $N<N_H$ number of supersymmetries. In particular,  we shall demonstrate for $H=SU(4)$ that  the group $\Sigma({\cal P}_{SU(4)})$
 can be used to find the Killing spinors for all backgrounds with $N<N_H=4$. It will become apparent
 that the  disconnected components of $\Sigma({\cal P}_{SU(4)})$ can also be used to reduce the number of choices of Killing spinors.

 The isotropy groups $H$ and the associated $\Sigma({\cal P}_H)$ groups of IIA and IIB supergravities have been tabulated
 in tables 4 and 6, respectively. These
 groups for IIB supergravity can easily be read off from those of heterotic supergravity, see also \cite{iib2}. For IIA
 supergravity, the isotropy groups $H$ and the associated $\Sigma({\cal P}_H)$ groups have a close relationship to those of eleven-dimensional
 supergravity. However, there are isotropy and $\Sigma({\cal P}_H)$ groups of the latter that do not have a IIA analogue.

This paper is organized as follows: In section two, we give the isotropy groups  and representatives
of the invariant spinors  of eleven-dimensional and type II supergravities. In section three, we give the
$\Sigma$ groups of eleven-dimensional and type II supergravity. In section four, we use the $\Sigma({\cal P}_{SU(4)})$
to give the normal forms of $SU(4)$-invariant Killing spinors, and emphasize the importance of the disconnected
components of the group, and in section five we give our conclusions. In the appendices A and B, we give details
of the computation of the isotropy and $\Sigma$ groups, respectively.

\newsection{Isotropy  groups and invariant spinors}

\subsection{Isotropy groups}

The gravitino and supersymmetry parameter  of eleven-dimensional supergravity are Majorana $Spin(10,1)$ spinors. To find the
isotropy groups\footnote{These are determined up to discrete identifications of the connected to the identity component.} of these spinors, one begins with the results of \cite{bryant, breakwave} which demonstrate that there are two kinds of orbits
of $Spin(10,1)$ on the space of Majorana spinors, $\Delta_{\bf 32}$, with isotropy groups $SU(5)$
and $Spin(7)\ltimes\bR^9$.
Then $\Delta_{\bf 32}$ is decomposed in irreducible representations of the above isotropy groups and the procedure is repeated until all
spinor singlets are found. This is essentially a Lie algebra computation and a more detailed description is given
in the appendices. The complete list of isotropy groups is given in table 1; for the spinor notation we use see \cite{ggp, syst}.
\begin{table}[tb]
 \begin{center}
\begin{tabular}{|c|c|}\hline
 $N_H$ & $H$
 \\ \hline \hline
 $1$& $Spin(7)\ltimes\bR^9$
 \\\hline
 $2$& $Spin(7), ~~SU(5), ~~SU(4)\ltimes\bR^9,~~ G_2\times\bR^9$
\\ \hline
$3$& $Sp(2)\times\bR^9$
\\ \hline
 $4$&  $SU(4),~~ G_2,~~ SU(2)\times SU(3), ~~(SU(2)^2)\ltimes \bR^9,~~ SU(3)\ltimes \bR^9$\\
\hline
 $5$&  $SU(2)\ltimes \bR^9$\\
\hline
$6$&  $Sp(2),~~ U(1) \ltimes \bR^9$\\
\hline
$8$&  $SU(3),~~ SU(2)^2,~~ SU(2)\ltimes \bR^9$ \\
\hline
$10$&  $SU(2)$\\
\hline
$12$&  $U(1)$\\
\hline
$16$&  $SU(2),~~ \bR^9$\\
\hline
$32$& $\{1\}$\\
\hline
\end{tabular}
\end{center}
\caption{The complete list of isotropy groups of spinors in eleven-dimensional supergravity. $H$ denotes the subgroups of $Spin(10,1)$,
up to discrete identifications of the connected to the identity component,
which leave some Majorana spinors invariant.  $N_H$ is the maximal number of $H$-invariant spinors. }
\end{table}

All non-compact isotropy groups are subgroups of $Spin(7)\ltimes\bR^9$. The same applies for the compact isotropy groups
apart from $SU(2)\times SU(3)$ and $G_2$. The former is subgroup only of $SU(5)$ while the latter is subgroup only of $Spin(7)\ltimes\bR^9$.
The rest of the compact isotropy groups are subgroups of both $SU(5)$ and $Spin(7)\ltimes \bR^9$. This easily follows from the analysis
of the singlets in the appendices. Many of the groups in table 1  and the cases
$H\subset Spin(7)\ltimes\bR^9$ have  previously appeared in \cite{breakwave, ggp} and in \cite{breakwave, null}, respectively.

In the non-compact cases, the group is a semi-direct product of a compact group $K$ and $\bR^9$. To specify the group, one has in addition to determine
the representation of  $K$  on $\bR^9$. This is easily found from the results of table 3 which give explicitly the invariant spinors.
In particular,  one has $(Spin(7)\ltimes \bR^8)\times\bR)$,  $(SU(4)\ltimes \bR^8)\times \bR$, $(G_2\ltimes\bR^7)\times \bR^2$, $(Sp(2)\ltimes \bR^8)\times\bR$,
$(SU(3)\ltimes \bR^6)\times \bR^3$,   $(SU(2)\ltimes (\oplus^2\bR^4))\times \bR$, $(SU(2)\ltimes\bR^4)\times (SU(2)\ltimes\bR^4)\times\bR$,
 $(U(1)\ltimes(\oplus^4\bR^2))\times\bR$ and $(SU(2)\ltimes \bR^4)\times\bR^5$. The groups are stated in the order in which they are denoted in table 1.

\subsection{Invariant spinors}

\subsubsection{Time-like basis}

To solve the Killing spinor equations, it is useful to have an explicit  basis in the
space of the $H$-invariant spinors. The most straightforward way to give such a basis is in terms
of the description of spinors in terms of forms. There are two ways
to describe the Majorana spinors, in terms of forms, associated with the construction of ${\rm Clif}(\bR^{10,1})$ from
either ${\rm Clif} (\bR^{10})$ or ${\rm Clif}(\bR^{9,1})$. These two constructions lead to different bases in ${\rm Clif}(\bR^{10,1})$, the ``time-like''
and ``null'' bases, respectively. These bases have been constructed  in  \cite{ggp,het, syst}, where one can also find the spinor conventions used in this paper.
The `time-like'' and ``null'' bases are suited to investigate backgrounds with $H\subseteq SU(5)$- and $H\subseteq Spin(7)\ltimes \bR^9$-invariant spinors,
respectively.  There are also several
cases which can be investigated using both kinds of bases. For future use, we shall give the invariant spinors in both
bases.

First, we consider the invariant spinors of $H\subseteq SU(5)$ isotropy groups in the {\it time-like basis} \cite{ggp, syst}.
The results of the detailed analysis in appendix A are presented in table 2.

\begin{table}[tb]
 \begin{center}
\begin{tabular}{|c|c|c|}\hline
 $N_H$ & $H\subseteq SU(5)$ &{\rm Singlets}
 \\ \hline \hline
 $2$& $SU(5)$ & $1$
\\ \hline
 $4$&  $SU(4)$& $1~,~e_5$\\
\hline
 $4$&  $SU(2)\times SU(3)$& $1~,~e_{12}$\\
\hline
$6$&  $Sp(2)$& $1~,~e_5~,~e_{12}+e_{34}$\\
\hline
$8$&  $SU(3)$& $1~,~e_5~,~e_4~,~e_{45}$\\
\hline
$8$&  $SU(2)^2$& $1~,~e_5~,~e_{12}~,~e_{34}$\\
\hline
$10$&  $SU(2)$& $1~,~e_5~,~e_{12}~,~e_{34}~,~e_{13}+e_{24}$\\
\hline
$12$&  $U(1)$& $1~,~e_5~,~e_{12}~,~e_{34}~,~e_{13}~,~e_{24}$\\
\hline
$16$&  $SU(2)$& $1~,~e_5~,~e_{12}~,~e_{34}~,~e_4~,~e_{45}~,~
e_3~,~e_{35}$\\
\hline
$32$&$\{1\}$& $\Delta_{\bf 32}$\\\hline
\end{tabular}
\end{center}
\caption{Explicit representatives of all $H$-invariant
spinors for all isotropy groups where $H\subseteq SU(5)$. If two singlets  are the real and imaginary parts of a complex (Dirac) spinor, then they have been given in terms of the
complex spinor. In particular,  the real and imaginary parts of the  the complex spinor $e_{i_1\dots i_k}$ are
$e_{i_1\dots i_k}+{(-1)^{[{k\over2}]}\over (5-k)!}  \epsilon_{i_1\dots i_k}{}^{j_{k+1}\dots j_5}
e_{j_{k+1}\dots j_5}$ and $i(e_{i_1\dots i_k}-{(-1)^{[{k\over2}]}\over (5-k)!}  \epsilon_{i_1\dots i_k}{}^{j_{k+1}\dots j_5}
e_{j_{k+1}\dots j_5})$ for $k=0,1,2$ and $j_1, \dots, j_k=1, \dots 5$, respectively \cite{ggp}.}
\end{table}


\subsubsection{Null basis}
The invariant spinors of $H\subseteq Spin(7)\ltimes \bR^9$ isotropy groups in the {\it null basis} \cite{syst} are
summarized  in table 3. A detailed derivation of these singlets can be found in appendix \ref{noncompsingletsapp}.
\begin{table}[tb]
 \begin{center}
\begin{tabular}{|c|c|c|}\hline
 $N_H$ & $H\subseteq Spin(7)\ltimes\bR^9$&{\rm Singlets}
 \\ \hline \hline
 $1$& $Spin(7)\ltimes \bR^9$ & $1+e_{1234}$
\\ \hline
$2$& $Spin(7)$ & $1+e_{1234}~,~e_5+e_{12345}$
\\
\hline
$2$&  $G_2\ltimes \bR^9$& $1+e_{1234}~,~e_{1}+e_{234}$
 \\
\hline
 $2$&  $SU(4)\ltimes\bR^9$& $1$
 \\
\hline
$3$&  $Sp(2)\ltimes\bR^9$& $1~,~i(e_{12}+e_{34})$
 \\
\hline
 $4$&  $SU(4)$& $1~,~e_5$
 \\
\hline
$4$&  $G_2$& $1+e_{1234}~,~~e_{15}+e_{2345}~,~~e_{1}+e_{234}~,~~e_5+e_{12345}$
 \\
\hline
$4$&  $(SU(2)^2)\ltimes\bR^9$& $1~,~e_{12}$
 \\
\hline
$4$&  $SU(3)\ltimes \bR^9$& $1~,~~e_{1}$
 \\
\hline
$5$&  $SU(2)\ltimes \bR^9$& $1~,~e_{12}~,~~e_{13}+e_{24}$
 \\
\hline
$6$&  $Sp(2)$& $1~,~i(e_{12}+e_{34})~,~e_5~,~~i(e_{125}+e_{345})$
 \\
\hline
$6$&  $U(1)\ltimes \bR^9$& $1~,~e_{12}~,~ e_{13}$
 \\
\hline
$8$&  $SU(2)^2$& $1~,~e_{12}~,~e_5~,~e_{125}$
 \\
\hline
$8$&  $SU(3)$& $1~,~e_{15}~,~e_5~,~e_{1}$
 \\
\hline
$8$& $SU(2)\ltimes \bR^9$& $1~,~e_{12}~,~e_{1}~,~e_2$\\
\hline
$10$&  $SU(2)$& $1~,~e_{12}~,~~e_{13}+e_{24}~,~e_5~,~e_{125}~,~e_{135}+e_{245}$
 \\
\hline
$12$&  $U(1)$& $1~,~e_{12}~,~e_{13}~,~e_5~,~e_{125}~,~e_{135}$
 \\
 \hline

$16$& $SU(2)$& $1~,~e_{12}~,~e_{15}~,~e_{25}~,~e_{1}~,~e_2~,~e_5~,~e_{125}$\\
\hline
$16$& $\bR^9$& $1~,~e_{12}~,~e_{13}~,~e_{14}~,~e_{1}~,~e_2~,~e_{3}~,~e_4$\\
\hline
$32$&$\{1\}$& $\Delta_{\bf 32}$\\\hline
\end{tabular}
\end{center}
\caption{Explicit representatives of all $H$-invariant
spinors for all isotropy groups where $H\subseteq Spin(7)\ltimes\bR^9$. If two singlets are the real and imaginary parts of a complex (Dirac) spinor, then they have been given in terms of the
complex spinor. In particular,  the real and imaginary parts of the  the complex spinor $e_{i_1\dots i_k}$  are
$e_{i_1\dots i_k}+{(-1)^{[{k\over2}]}\over (4-k)!} \epsilon_{i_1\dots i_k}{}^{j_1\dots j_{4-k}} e_{j_1\dots j_{4-k}}$
and $i(e_{i_1\dots i_k}-{(-1)^{[{k\over2}]}\over (4-k)!} \epsilon_{i_1\dots i_k}{}^{j_1\dots j_{4-k}} e_{j_1\dots j_{4-k}})$, respectively, for $k=0,1,2$
and $i,j=1,\dots, 4$. In addition, $e_{i_1\dots i_k 5}=({\rm Re}\,e_{i_1\dots i_k})\wedge e_5+ i ({\rm Im}\,e_{i_1\dots i_k}) \wedge e_5$ \cite{syst}.  }
\end{table}

A consequence of the results summarized in tables 2 and 3 is that  there are restrictions on the number of singlets that can occur.
In particular,  the isotropy group
of more than 16 linearly independent spinors is the identity. In addition, there are no cases with $N_H=7, 11, 13, 14$ and $15$.
Furthermore, for every non-compact isotropy group there is an associated compact one with twice the number of invariant Killing spinors.

\subsection{Comparison with the isotropy groups of IIB and IIA supergravities}

The relevant spinor representation of IIB supergravity is the complex chiral (Weyl) representation ${}^c\Delta^{+}_{\bf 16}$ of $Spin_c(9,1)$.
This is constructed from the associated Majorana-Weyl representation, $\Delta^+_{\bf 16}$, associated with the heterotic string, by a straightforward
complexification, ${}^c\Delta^{+}_{\bf 16}=\Delta^+_{\bf 16}\otimes\bC=\Delta^+_{\bf 16}\oplus i \Delta^+_{\bf 16}$ . As a result the isotropy groups for spinors of  IIB supergravity in $Spin_c(9,1)$ are precisely those
found for the heterotic string. The associated invariant spinors  are the complexification of those of heterotic supergravity.
One consequence of this is that in IIB supergravity there are always an even number of invariant spinors.
For example in heterotic supergravity the spinor $1+e_{1234}$ is $Spin(7)\ltimes\bR^8$ invariant. The associated invariant spinors
of IIB supergravity are $1+e_{1234}$ and $i(1+e_{1234})$.

The relevant spinor representation of IIA supergravity is the Majorana representation, $\Delta_{\bf 32}$, of $Spin(9,1)$. This should be thought
of as the direct sum of the positive and negative chirality Majorana-Weyl representations of $Spin(9,1)$,
$\Delta_{\bf 32}=\Delta_{\bf 16}^+\oplus \Delta^-_{\bf 16}$. Since a single spinor
in a Majorana-Weyl representation of $Spin(9,1)$ has isotropy group $Spin(7)\ltimes \bR^8$, it is clear that all the isotropy
groups of spinors of IIA supergravity must be subgroups of $Spin(7)\ltimes \bR^8$. It is then straightforward to show
that the isotropy groups are closely related to those of table 3 for eleven-dimensional supergravity. If the isotropy
group in table 3 is of the type $K\ltimes \bR^9$, then the associated isotropy group in IIA supergravity is $K\ltimes\bR^8$.
The additional generator  is along the additional 11th direction.
Moreover the invariant spinors are precisely as those given in table 3 now interpreted as $Spin(9,1)$ spinors. This follows from the well-known fact that
the Majorana representation of $Spin(10,1)$ decomposes under $Spin(9,1)$ as the sum of a positive and negative
chirality Majorana-Weyl representation. The isotropy groups, up to discrete identifications of the connected to the identity component,
of IIA and IIB supergravity spinors
 as well as the singlets in the null basis \cite{het} are summarized in
table 4.


\begin{table}[h!]
 \begin{center}
\begin{tabular}{|c|c|c|c|}\hline
 $H\subseteq Spin(7)\ltimes\bR^8$&${\rm IIA}/ N_H$&${\rm IIB}/ N_H$&{\rm Singlets}
 \\ \hline \hline
 $Spin(7)\ltimes \bR^8$ &$1$&$2$& $1+e_{1234}$
\\ \hline
 $Spin(7)$&$2$&$-$ & $1+e_{1234}~,~e_5+e_{12345}$
\\
\hline
$G_2\ltimes \bR^8$&$2$&$-$& $1+e_{1234}~,~e_{1}+e_{234}$
 \\
\hline
 $SU(4)\ltimes\bR^8$&$2$&$4$& $1$
 \\
\hline
 $Sp(2)\ltimes\bR^8$&$3$&$6$& $1~,~i(e_{12}+e_{34})$
 \\
\hline
 $SU(4)$&$4$&$-$& $1~,~e_5$
 \\
\hline
$G_2$&$4$&$4$& $1+e_{1234}~,~~e_{15}+e_{2345}~,~~e_{1}+e_{234}~,~~e_5+e_{12345}$
 \\
\hline
$(SU(2)^2)\ltimes\bR^8$&$4$&$8$& $1~,~e_{12}$
 \\
\hline
 $SU(3)\ltimes \bR^8$&$4$&$-$& $1~,~~e_{1}$
 \\
\hline
$SU(2)\ltimes \bR^8$&$5$&$10$& $1~,~e_{12}~,~~e_{13}+e_{24}$
 \\
\hline
$Sp(2)$&$6$&$-$& $1~,~i(e_{12}+e_{34})~,~e_5~,~~i(e_{125}+e_{345})$
 \\
\hline
$U(1)\ltimes \bR^8$&$6$&$12$& $1~,~e_{12}~,~ e_{13}$
 \\
\hline
$SU(2)^2$&$8$&$-$& $1~,~e_{12}~,~e_5~,~e_{125}$
 \\
\hline
 $SU(3)$&$8$&$8$& $1~,~e_{15}~,~e_5~,~e_{1}$
 \\
\hline
$SU(2)\ltimes \bR^8$&$8$&$-$& $1~,~e_{12}~,~e_{1}~,~e_2$\\
\hline
 $SU(2)$&$10$&$-$& $1~,~e_{12}~,~~e_{13}+e_{24}~,~e_5~,~e_{125}~,~e_{135}+e_{245}$
 \\
\hline
  $U(1)$&$12$&$-$& $1~,~e_{12}~,~e_{13}~,~e_5~,~e_{125}~,~e_{135}$
 \\
 \hline
 $SU(2)$&$16$&$16$& $1~,~e_{12}~,~e_{15}~,~e_{25}~,~e_{1}~,~e_2~,~e_5~,~e_{125}$\\
\hline
 $\bR^8$&$16$&$16$& $1~,~e_{12}~,~e_{13}~,~e_{14}~,~e_{1}~,~e_2~,~e_{3}~,~e_4$\\
\hline
$\{1\}$&$32$&$32$& ${}^c\Delta^{+}_{\bf 16}~,~\Delta^{+}_{\bf 16}\oplus \Delta^{-}_{\bf 16}$\\\hline
\end{tabular}
\end{center}
\caption{The complete list of isotropy groups, up to discrete identifications of the connected to the identity component, of IIA and IIB supergravity spinors
 including explicit representatives for the singlets in the null basis. $N_H$ denotes the real dimension of ${\cal P}_H$ in the IIA and IIB cases, and $-$ denotes cases that do not occur
in IIB. The singlets are given for both IIA and IIB supergravities, and the reality conditions are as those given
in table 3. The $H$-invariant IIB spinors
are always even forms, while the spinors of IIA are both even and odd forms.
For example, the $SU(4)\ltimes \bR^8$-invariant spinors in IIB are $1$ and $i1$ or, equivalently, $1+e_{1234}, i(1-e_{1234}), i(1+e_{1234})$ and $1-e_{1234}$ over the reals in the Majorana-Weyl basis.}
\end{table}

It is clear that all the isotropy groups of IIB and IIA supergravities appear as subgroups of the isotropy groups of spinors of eleven-dimensional
supergravity but the converse is not true.
In particular, there are isotropy groups of eleven-dimensional supergravity that do not have a type II analogue.
These are the  $SU(5)$ and $SU(2)\times SU(3)$ isotropy groups. Clearly, supersymmetric backgrounds with such invariant
Killing spinors have a purely eleven-dimensional origin.
\newsection{The $\Sigma({\cal P}_H)$ groups}

\subsection{Eleven-dimensional supergravity}

For generic backgrounds the holonomy of the supercovariant connection of eleven-dimen\-sional supergravity
is $SL(32,\bR)$  \cite{hull, duff, tsimpis}.
A consequence of this, in the context of backgrounds with $H$-invariant Killing spinors, is that in most cases one expects
that there are solutions
with any number $N$ of supersymmetries for $N\leq N_H$, where $N_H={\rm dim}\, \Sigma({\cal P}_H)$.
Although this is the expectation, there are also exceptions
  \cite{iibm31, bandos}, and a conjecture for the fractions that can occur can be found in \cite{duffb}.

Assuming that the Killing spinors $\e$ of a
 supersymmetric background are $H$-invariant, $N\leq N_H$, these can be written as a linear combination
of a basis $(\eta_i)$, $i=1,\dots, N_H$,  in tables 2 or 3, i.e.
\bea
\e=\sum_{i=1}^{N_H} f_i \eta_i~,
\eea
where $f_i$ are spacetime functions.
To solve the Killing spinor equation, it is convenient to bring the Killing spinors $\e$ to a normal form.
To do  this, one may consider the subgroup of  $Spin(10,1)$ which leaves the plane ${\cal P}_H$ of {\it all} $H$-invariant spinors
invariant. This  is clearly the remaining gauge group of the theory. Since the isotropy group $H$ acts on ${\cal P}_H$
with the identity transformation, it is clear that we should consider those transformations of $Spin(10,1)$ which leave
${\cal P}_H$ invariant up to transformations generated by $H$. This  is precisely the $\Sigma({\cal P}_H)$ group for the subspace ${\cal P}_H$ using the
definition of \cite{hetb}.

\begin{table}[h]
 \begin{center}
\begin{tabular}{|c|c|c|}\hline
 $N_H$ & $H$ & $\Sigma({\cal P}_H)$
 \\ \hline
 \hline
 $1$& $Spin(7)\ltimes\bR^9$& $Spin(1,1)$
 \\\hline
 $2$& $Spin(7)$&$Spin(2,1)$
 \\ \hline
 &$SU(5)$& $U(1)$
 \\ \hline
 &$SU(4)\ltimes\bR^9$&$Spin(1,1)\times U(1)$
 \\ \hline
 &$G_2\ltimes\bR^9$&$Spin(1,1)\times U(1)$
\\ \hline
$3$& $Sp(2)\ltimes\bR^9$&$Spin(1,1)\times SU(2)$
\\ \hline
 $4$&  $SU(4)$& $Spin(2,1)\times U(1)$
 \\ \hline
 &$G_2$&$Spin(3,1)$
 \\ \hline
 &$SU(2)\times SU(3)$& $SU(2)\times U(1)$
 \\ \hline
 &$(SU(2)^2)\ltimes \bR^9$&$Spin(1,1)\times SU(2)^2$
 \\ \hline
 &$ SU(3)\ltimes \bR^9$&$Spin(1,1)\times SU(2)\times U(1)$
 \\
\hline
 $5$&  $SU(2)\ltimes \bR^9$&$Spin(1,1)\times Sp(2)$
 \\
\hline
$6$&  $Sp(2)$& $Spin(2,1)\times SU(2)$
\\
\hline
&$ U(1) \ltimes \bR^9$&$Spin(1,1)\times SU(4)$
\\
\hline
$8$&  $SU(3)$& $Spin(4,1)\times U(1)$
\\
\hline
&$ SU(2)^2$& $Spin(2,1)\times SU(2)^2$
\\
\hline
&$ SU(2)\ltimes \bR^9$&$Spin(1,1)\times Sp(2)\times SU(2)$
 \\
\hline
$10$&  $SU(2)$&$Spin(2,1)\times Sp(2)$\\
\hline
$12$&  $U(1)$&$Spin(2,1)\times SU(4)$\\
\hline
$16$&  $SU(2)$&$Spin(6,1)\times SU(2)$\\
\hline
&$\bR^9$&$Spin(1,1)\times Spin(9)$
\\
\hline
$32$& $\{1\}$&$Spin(10,1)$\\
\hline
\end{tabular}
\end{center}
\caption{The $\Sigma({\cal P}_H$) groups for each possible isotropy group of spinors in eleven-dimensional supergravity. The explicit action of the generators of $\Sigma({\cal P}_H$) on ${\cal P}_H$ can be found in appendix
B.}
\end{table}

The computation of the $\Sigma({\cal P}_H)$ groups for each $H$ can be done as in heterotic supergravity (type I)  \cite{hetb}.
The details can be found in the appendices. The results are given in table 5.

It is clear from the results of table 5 that the $\Sigma$ groups are a product, $Spin\times R$, where $Spin$
and $R$ are the $Spin$ group and the $R$-symmetry group of a lower-dimensional supergravity theory.
This allows us to view the associated eleven-dimensional supersymmetric backgrounds as being in the same universality class
as those lifted from the lower-dimensional supergravity theories constructed from compactification on a holonomy $H$ manifold.
However, this does not imply that the  associated backgrounds  have a lower dimensional origin.
In two cases,
those with isotropy groups $SU(5)$ and $SU(2)\times SU(3)$, the associated $Spin$ group is the identity. This is because
such eleven-dimensional backgrounds are in the same universality class as those associated with compactifications
of eleven-dimensional supergravity on holonomy $SU(5)$ and $SU(2)\times SU(3)$ manifolds to one dimension, and the $Spin$
group in one dimension, up to a discrete identification, is the identity.

It is also important to notice that the $\Sigma({\cal P}_H)$ groups do not capture the full expected $R$-symmetry groups of the
associated lower dimensional supergravity theories. For example consider the $N_H=8$,  $H=SU(3)$ case
in table 5. The associated lower dimensional theory is a 5-dimensional supergravity with 8 real supersymmetries and the
total $Spin$ and $R$-symmetry group is expected to be $Spin(4,1)\times SU(2)$. However
$\Sigma({\cal P}_H)=Spin(4,1)\times U(1)\subset Spin(4,1)\times SU(2)$. This is because, unlike the $R$-symmetry groups
of lower-dimensional supergravities in general,
 the $R$-symmetry subgroups that appear in
 $\Sigma({\cal P}_H)$ are required to be subgroups of $Spin(10,1)$.
 As a result, in some cases the the $R$-symmetry group contained in $\Sigma({\cal P}_H)$ is only a subgroup of the
 $R$-symmetry group of the lower-dimensional supergravity theory.

\subsection{IIA and IIB supergravities}

For completeness, we also give the $\Sigma$ groups of the invariant spinors of type II supergravities.
The $\Sigma$ groups of type IIB supergravity can easily be derived from those of type I supergravity.
One difference is that there is an additional $U(1)$ generator because of  the $Spin_c$ nature
of IIB spinors. The $\Sigma$ groups of IIA supergravity can easily be derived from those of eleven-dimensional
supergravity. In particular, one simply excludes all the generators associated with the eleventh direction.
The results are tabulated in table 6.

\begin{table}[h]
 \begin{center}
\begin{tabular}{|c|c|c|}\hline
 $H$ & $\Sigma({\cal P}_H)/IIA ~(N_H)$& $\Sigma({\cal P}_H)/IIB~(N_H)$
 \\ \hline
 \hline
$Spin(7)\ltimes\bR^8$&$Spin(1,1)~(1)$&$Spin_c(1,1)~(2)$
 \\\hline
  $Spin(7)$&$Spin(1,1)~ (2)$&$-$
 \\ \hline
 $SU(4)\ltimes\bR^8$&$Spin(1,1)\times U(1)~(2)$&$Spin_c(1,1)\times U(1)~ (4)$
 \\ \hline
 $G_2\ltimes\bR^8$&$Spin(1,1)~(2)$&$-$
\\ \hline
$Sp(2)\ltimes\bR^8$&$ Spin(1,1)\times SU(2)~(3)$&$Spin_c(1,1)\times SU(2)~(6)$
\\ \hline
 $SU(4)$& $Spin(1,1)\times U(1)~(4)$&$-$
 \\ \hline
 $G_2$&$Spin(2,1)~(4)$&$Spin_c(2,1)~(4)$
 \\ \hline
 $(SU(2)^2)\ltimes \bR^8$&$Spin(1,1)\times SU(2)^2~(4)$&$Spin_c(1,1)\times SU(2)^2~(8)$
 \\ \hline
 $ SU(3)\ltimes \bR^8$&$ Spin(1,1)\times U(1)^2~(4)$&$-$
 \\
\hline
  $SU(2)\ltimes \bR^8$&$Spin(1,1)\times Sp(2)~(5)$&$Spin_c(1,1)\times Sp(2)~(10)$
 \\
\hline
 $Sp(2)$& $Spin(1,1)\times SU(2)~(6)$&$-$
\\
\hline
$ U(1) \ltimes \bR^8$&$ Spin(1,1)\times SU(4)~(6)$&$Spin_c(1,1)\times SU(4)~(12)$
\\
\hline
 $SU(3)$& $Spin(3,1)\times U(1)~(8)$&$Spin_c(3,1)\times U(1)~(8)$
\\
\hline
$ SU(2)^2$& $Spin(1,1)\times SU(2)^2~(8)$&$-$
\\
\hline
$ SU(2)\ltimes \bR^8$&$Spin(1,1)\times Spin(4)\times SU(2)~(8)$&$-$
 \\
\hline
$SU(2)$&$Spin(1,1)\times Sp(2)~(10)$&$-$\\
\hline
$U(1)$&$Spin(1,1)\times SU(4)~(12)$&$-$\\
\hline
$SU(2)$&$Spin(5,1)\times SU(2)~(16)$&$Spin_c(5,1)\times SU(2)~(16)$\\
\hline
$\bR^8$&$Spin(1,1)\times Spin(8)~(16)$&$Spin_c(1,1)\times Spin(8)~(16)$
\\
\hline
$\{1\}$&$Spin(9,1)~(32)$&$Spin_c(9,1)~(32)$\\
\hline
\end{tabular}
\end{center}
\caption{The $\Sigma({\cal P}_H$) groups for each possible isotropy group of spinors in IIA and IIB supergravity.  The first column contains the isotropy groups of spinors of type II supergravities. The second column denotes
the $\Sigma$ groups of IIA supergravity. The number in $(\cdot)$ denote the real dimension of ${\cal P}_H$. The third column
contains the $\Sigma$ groups of IIB supergravity and $-$
denotes the IIB cases that do not occur.  }
\end{table}

\newsection{Disconnected components of $\Sigma$ groups}

The Lie algebra computation that we have done, which is summarized in the appendices,
identifies the component of each $\Sigma$ group that is connected to the identity. However, the  $\Sigma$ groups may have disconnected
components. Since these   disconnected components of the $\Sigma$
groups were not essential for the solution of Killing spinor equations of heterotic supergravity, they were not stressed
in that computation. This is no longer the case for type II and eleven-dimensional supergravities.

\subsection{Discrete subgroups}

We shall not attempt to compute the disconnected components of all $\Sigma$ groups. Instead we shall focus on the case of $SU(4)$
invariant spinors. Before we proceed with this, let us establish some notation. It is known that field theories
with chiral couplings may violate parity invariance. So the minimal requirement
imposed on a relativistic theory is that it should be covariant under {\it restricted} Lorentz transformations, i.e.~the transformations of the connected component of the Lorentz group. These transformations can also be characterized as
{\it proper} and {\it orthochronous}, i.e.~those that preserve both the orientation of spacetime and the direction of time. The spin group  associated to the restricted Lorentz transformations is the connected component $Spin^0$ of $Spin$; $Spin$ is the
double cover of the proper Lorentz transformations.

Suppose that we consider those $\Sigma$ groups that are constructed from $Spin^0(n, 1)$ transformations. It is expected that in such a case
$\Sigma= Spin^0(d,1)\times R$,
where $d<n$. Although $Spin^0(d,1)$ is  connected, we shall see that $R$ can be disconnected.
 For this let us consider the case of eleven-dimensional supergravity
backgrounds with $SU(4)$-invariant spinors. In this case, the connected component $\Sigma^0$ of the $\Sigma$ group,
up to discrete identifications that preserve  component connected to the identity,
 is $\Sigma^0({\cal P}_{SU(4)})=Spin^0(2,1)\times U(1)$, see appendix B. These are not
the only transformations of $Spin^0(10,1)$ that preserve ${\cal P}_{SU(4)}$. It is easy to see that the discrete $Spin^0(10,1)$ transformations
\bea
&&\Gamma_{1234}~,~~~\Gamma_{6234}~,~~~\Gamma_{1734}~,~~~\Gamma_{1284}~,~~~\Gamma_{1239}~,~~~\Gamma_{6734}~,~~~\Gamma_{6284}~,~~~
\cr
&&\Gamma_{6239}~,~~~\Gamma_{1784}~,~~~\Gamma_{1739}~,~~~\Gamma_{1289}~,~~~\Gamma_{1789}~,~~~\Gamma_{6289}~,~~~\Gamma_{6739}~,~~~
\cr
&&\Gamma_{6784}~,~~~\Gamma_{6789}~,
\la{dis}
\eea
also leave  ${\cal P}_{SU(4)}$ invariant and are not in $Spin^0(2,1)\times U(1)$.

\subsection{Killing spinors}

To illustrate the importance of the additional transformations in $\Sigma({\cal P}_{SU(4)})$,
let us investigate the orbits of $Spin^0(2,1)\times U(1)$
on ${\cal P}_{SU(4)}$. It turns out that there are three types of orbits with representatives
\bea
1+e_{12345}~,~~~e_5+e_{1234}~,~~~1+e_{12345}+e_5+e_{1234}~.
\eea
The first two represent orbits with compact isotropy groups and  the last represents an orbit with a non-compact isotropy group.
This means that there are three cases that should be investigated for backgrounds with $N=1$ supersymmetry.
This appears to be a contradiction because  it is known that there are only two types of orbits of $Spin^0(10,1)$ on the space of
Majorana spinors with isotropy groups
$SU(5)$ and $Spin(7)\ltimes\bR^9$ \cite{bryant, breakwave}. Of course this may
imply that the transformation of $Spin^0(10,1)$ which
relates the first two orbits
does not preserve ${\cal P}_{SU(4)}$. However, this is not the case. Observe that the first element in (\ref{dis}) transforms the representative
of the first orbit to the second. This in turn gives only two distinct cases with $N=1$ supersymmetry. This may not seem
to be advantageous since for selecting the first Killing spinor the whole $Spin^0(10,1)$ could be used. However, it has
an effect when investigating cases with $N=3$ supersymmetry where only the $\Sigma({\cal P}_{SU(4)})$ group can be used.
So there are two cases with $N=3$ supersymmetry and $SU(4)$-invariant spinors that should be investigated instead of the three which
arise from considering only the connected component of the ${\cal P}_{SU(4)}$ group.

Of course, the representatives  $1+e_{12345}$ and $e_5+e_{1234}$ of the orbits can also be related by the $\Gamma_{0\nat}$, $\nat=10$,
transformation written in the timelike basis, which is in $Spin(2,1)$. However $\Gamma_{0\nat}$ is not in either $Spin^0(10,1)$ or in $Spin^0(2,1)$
because it induces non-orthochronous Lorentz transformations. So if the eleven-dimensional theory is only assumed to be
invariant under restricted Lorentz transformations, then the only way to relate the two compact orbits is with
a discrete $R$ transformation in (\ref{dis}). The analysis presented above can easily be be modified if one begins
with $Spin(10,1)$ rather than $Spin^0(10,1)$.

\subsection{Compactifications}

To investigate the role of the discrete transformations in the context of compactifications, consider an $\bR^{2,1}\times X$ compactification
with four  $SU(4)$-invariant Killing spinors, where $X$
is an $SU(4)$-structure manifold. Again we assume that the eleven-dimensional theory we compactify has local $Spin^0(10,1)$
invariance. The vacuum configuration breaks this to the $\Sigma({\cal P}_{SU(4)})=Spin^0(2,1)\times R$ group.

To separate the $Spin^0(2,1)$ transformations from those of $R$, it suffices to see how they act on the frame of
$\bR^{2,1}\times X$.
The frame of the internal space $X$ is $(e^a, e^{5+a})$, $a=1,2,3,4$,
and the remaining three directions
are those of $\bR^{2,1}$. It is easy to see that $Spin^0(2,1)$ acts with restricted Lorentz transformations on the
frame of $\bR^{2,1}$ and leaves invariant the frame of the internal space, while the $R$ transformations
leave the frame of $\bR^{2,1}$ invariant while transforming the frame of the internal space.

To see the transformations induced by the discrete transformations (\ref{dis}), observe that $\Gamma_{1234}$ acts on $e^a$ as
\bea
e^a\rightarrow -e^a~,~~~a=1,2,3,4,
\la{disa}
\eea
leaving the rest of the frame directions invariant.  It is straightforward to compute the action of the rest of the elements in (\ref{dis}).
Such transformations leave the metric of the internal space invariant but act on the fluxes. Moreover they act on the spinors
of the theory and change the (almost)
complex structure $I$ of the internal space to $-I$.

The action on the fluxes can be identified by observing that $\Gamma_{1234}$  changes the  $(p,q)$ tensors, with respect to $I$,
to $(q,p)$ ones. The action on the fermions is straightforward since $\Gamma_{1234}$ is an element of the $Spin^0(10,1)$ group.
Since from the perspective of the lower-dimensional theory (\ref{disa}) is a remnant of the {\it restricted} Lorentz
group in eleven dimensions, it is natural to argue that it must remain  a symmetry after compactification.
It is clear that such an assertion
will put restrictions on the couplings of the lower-dimensional effective theory.

The action of $\Gamma_{1234}$ on the fluxes and spinors is a $\bZ_2$ action. As such it resembles an $R$-parity transformation
\cite{rparity}. Though in our case the $\Sigma$ group has many disconnected components, so the representatives  (\ref{dis}) form a larger
group of reflections.

\newsection{Concluding Remarks}

We have given the isotropy groups of spinors of eleven-dimensional and type II supergravities as well as representatives
of the singlets. Using these, we have computed the $\Sigma({\cal P}_H)$ group of  ${\cal P}_H$, where ${\cal P}_H$ is the plane spanned
by {\it all}  $H$-invariant spinors. These are the subgroups of the gauge groups of ten- and eleven-dimensional supergravities
which preserve ${\cal P}_H$.  We have found
that the $\Sigma({\cal P}_H)$ groups are of the type $Spin\times R$, i.e.~they are the product of a $Spin$ group and an $R$-symmetry group of a lower-dimensional
supergravity. In some cases though the $R$-symmetry group contained in $\Sigma({\cal P}_H)$ is a subgroup of the $R$-symmetry
group of the associated lower-dimensional supergravity. This is because not all $R$-symmetry groups of lower-dimensional supergravities
are subgroups of either $Spin(9,1)$ or $Spin(10,1)$.

The solution of the Killing spinor equations of backgrounds with $H$-invariant spinors  can
 proceed as described in \cite{ggp} and \cite{iibm31}. In particular, if
$N$ is  small, then $\Sigma({\cal P}_H)$ can be used to find simplified canonical forms for the Killing spinors. 
On the other hand, if $N$ is near $N_H$, then $\Sigma({\cal P}_H)$ can be used to find the canonical form of the
normals to the Killing spinors in ${\cal P}_H$ or its dual.
In addition, we have emphasized the role of the disconnected components of the  $\Sigma({\cal P}_H)$ groups in choosing
 representatives for the Killing spinors.
It is expected that in this way
one can solve the Killing spinor equations of eleven-dimensional and type II supergravities
for all $H$-invariant Killing spinors.

As we have mentioned the disconnected components of the $\Sigma({\cal P}_H)$ groups in the context of compactifications should lead to
discrete symmetries
in the lower-dimensional supergravity effective theories. These are reminiscent of $R$-parity type of transformations, see e.g.~\cite{rparity}.
 In the $SU(4)$ case that we have investigated in some detail, the discrete group has several generators which act
 like reflections on the frame, on the fluxes and on the fermions of the theory. In a consistently
 constructed effective theory for a compactification, the invariance under such discrete $R$-symmetries may be manifest.
 This is because they are remnants of the Lorentz symmetry in higher dimensions. Nevertheless in the absence of a constructive
 method for specifying an effective theory, they may provide additional symmetry information which may lead to the suppression
 of some couplings.

\vskip 0.5cm
\noindent{\bf\large Acknowledgements} \vskip 0.3cm

The authors would like to thank Daniel Persson,
Axel Kleinschmidt and Antoine Van Proeyen for discussions. 
The work of UG is funded by the Swedish Research Council.

\vskip 0.5cm

\setcounter{section}{0}

\appendix {Invariant spinors}

\setcounter{subsection}{0}
\subsection{The $SU(5)$ series}

There are two orbits of $Spin(10,1)$ on the Majorana representation $\Delta_{\bf 32}$ of $Spin(10,1)$ with isotropy groups
$SU(5)$ and $Spin(7)\ltimes\bR^9$. First consider the $SU(5)$ case. To simplify notation, let us use standard notation and denote the representations by the
dimension.
 The 32-dimensional Majorana $Spin(10,1)$
representation decomposes under $SU(5)$ as

 \bea
  {\bf
32}= {\bf 1}+ \bar {\bf 1}+ {\bf 5}+\bar {\bf 5}+ {\bf
10}+\bar {\bf 10}~.
\eea
The singlets can be arranged  in the directions
\bea
1+e_{12345}~,~~~i(1-e_{12345})~.
\la{ssu5}
\eea

To find all the singlets and isotropy groups, we shall do the computations in several steps.

{\it Step 1}:
An additional  singlet can be either in
the ${\bf 5}$
or in ${\bf 10}$ representations.
If it is in the ${\bf 5}$ representation  the isotropy group is
$SU(4)$. Moreover under $SU(4)$ one has that
\bea
{\bf 32}=
+^2{\bf 1}+^2 \bar {\bf 1}+^2 {\bf 4}+^2\bar {\bf
4}+ {\bf 6}+\bar {\bf 6}~.
\la{su4}
\eea
The additional $SU(4)$-invariant spinors are
\bea
e_5+e_{1234}~,~~~i(e_5-e_{1234})~.
\la{ssu4}
\eea

On the other hand if the additional singlet  is in ${\bf 10}$, there are three possibilities because there are three kinds of orbits \cite{ggp}.
One is the generic
orbit with isotropy group $SU(2)^2$, and two special orbits with stability subgroups $Sp(2)$ and $SU(2)\times SU(3)$, respectively.
Under $SU(2)^2$, we have the decomposition
\bea
{\bf 32}=+^4
{\bf 1}+^4 \bar {\bf 1}+^2 ({\bf 2}, {\bf 2})+^2(\bar{\bf 2}, \bar{\bf 2})
+{\bf 2}\otimes {\bf 2}+\bar{\bf 2}\otimes \bar{\bf 2}~.
\la{su2su2}
\eea
The additional singlets to those of (\ref{ssu5}) are
\bea
&&e_5+e_{1234}~,~~~i(e_5-e_{1234})~,~~~e_{12}-e_{345}~,~~~
\cr
&&i(e_{12}+e_{345})~,~~~e_{34}-e_{125}~,~~~i(e_{34}+e_{125})~.
\la{ssu2su2}
\eea

Decomposing the spinors under $Sp(2)$, we have
\bea
{\bf 32}=+^3 {\bf 1}+^3 \bar {\bf 1}+^2 {\bf 5}+^2 {\bf 8}~.
\la{sp2}
\eea
The additional singlets to those of (\ref{ssu5}) are
\bea
e_5+e_{1234}~,~~~i(e_5-e_{1234})~,~~~e_{12}+e_{34}-e_{345}-e_{125}~,~~~i(e_{12}+e_{34}+e_{345}+e_{125})~.
\eea

Considering the isotropy group $SU(2)\times SU(3)$, the decomposition reads
\bea
{\bf 32}=
+^2 {\bf 1}+^2 \bar {\bf 1}+ ({\bf 2}, {\bf 3})+(\bar{\bf 2}, \bar{\bf 3})+{\bf 2}\otimes {\bf 3}+
\bar{\bf 2}\otimes \bar{\bf 3}+({\bf 1}, {\bf 3})+({\bf 1}, \bar{\bf 3})~.
\la{su2su3}
\eea
The additional singlets to those of (\ref{ssu5}) are
\bea
e_{12}-e_{345}~,~~~i(e_{12}+e_{345})~.
\la{ssu2su3}
\eea

{\it Step 2-1}:  Let us begin with (\ref{su4}). The singlets can be
either in the ${\bf 4}$ or the ${\bf 6}$ representation. If the
additional singlets are in ${\bf 4}$, then the isotropy group is $SU(3)$. In addition the Majorana representation decomposes under $SU(3)$ as
\bea
{\bf 32}=
+^4 {\bf 1}+^4 \bar {\bf 1}+^4 {\bf 3}+^4\bar {\bf
3}~.
\la{su3}
\eea
The additional singlets to those of (\ref{ssu4}) can be chosen as
\bea
e_4-e_{1235}~,~~~i (e_4+e_{1234})~,~~~e_{45}-e_{123}~,~~~i(e_{45}+e_{123})~.~~~
\la{ssu3}
\eea
Now if the additional singlet is in the ${\bf 6}$ representation, there are different isotropy groups that can occur
depending on the choice of orbit. These are two different orbits with isotropy groups $SU(2)^2$ and $Sp(2)$, respectively.
The decomposition according the former
is as in (\ref{su2su2}) while for the latter is as in (\ref{sp2}).

{\it Step 2-2}: The additional singlet in (\ref{su2su2}) can be either in the $({\bf 2}, {\bf 2})$ or the ${\bf 2}\otimes {\bf 2}$
representations. In the former case the isotropy group is $SU(2)$ and the decomposition is
\bea
{\bf 32}=+^8 {\bf 1}+^8 \bar{\bf 1}\oplus^4 {\bf 2}\oplus^4 \bar{\bf 2}~.
\la{su2}
\eea
This is the standard $SU(2)$ case which preserves 16 supersymmetries. The additional singlets to those of (\ref{ssu5}) and (\ref{ssu4}) are
\bea
&&e_4-e_{1235}~,~~~i (e_4+e_{1234})~,~~~e_{45}-e_{123}~,~~~i(e_{45}+e_{123})~,
\cr
&&e_3+e_{1245}~,~~~i(e_3-e_{1245})~,~~~
e_{35}+e_{124}~,~~~i(e_{35}-e_{124})~.
\la{ssu2}
\eea

If on the other hand the singlet is in ${\bf 2}\otimes {\bf 2}$, there are two orbits to consider one with isotropy group $SU(2)$ and
the other with isotropy group $U(1)$. In the $SU(2)$ case the decomposition is
\bea
{\bf 32}\rightarrow
+^5{\bf 1}+^5 \bar {\bf 1}+^4{\bf 2}+^4\bar{\bf 2}+^2{\bf 3}~.
\la{so3}
\eea
The additional singlets to those of (\ref{ssu5}) and (\ref{ssu2su2})   are
\bea
e_{13} +e_{24}+e_{245}+e_{135}~,~~~i(e_{13}+e_{24}-e_{245}-e_{135})~.
\la{sso3}
\eea
For the $U(1)$ orbit, the decomposition is
\bea
{\bf 32}=
+^6{\bf 1}+^6 \bar {\bf 1}+^{10}{\bf 2}~.
\la{u1}
\eea
The additional singlets to those of (\ref{ssu5}) and (\ref{ssu2su2})   are
\bea
e_{13} +e_{245}~,~~~i(e_{13}-e_{245})~,~~~ e_{24}+e_{135}~,~~~i(e_{24}-e_{135})~.
\la{ssu1}
\eea

{\it Step 2-3}:  Next consider the $Sp(2)$ case. The additional singlets can be either in the ${\bf 5}$ representation or in the ${\bf 8}$
representation. In the former case, it reduces to the $SU(2)^2$ case (\ref{su2su2}), and in the latter it reduces to $SU(2)$ (\ref{su2}).

{\it Step 2-4}:   The additional singlets can either be in the $({\bf 2},{\bf 3})$ or in the ${\bf 2}\otimes {\bf 3}$
representations. The former case reduces to either (\ref{su3}) or (\ref{su2su2}) cases. In the latter case,
it reduces to the (\ref{su2su2}) case and its descendants\footnote{Observe that $SU(2)\times SU(3)$ has two
kinds of orbits on ${\bf 2}\times {\bf 3}$ represented by the rank 1 and rank 2 $2\times 3$ matrices. The latter case reduces
to (\ref{su2su2}) and the former to the $U(1)$ case.}.

{\it Step 3-1}: Consider first the $SU(3)$ case. The additional singlet can be in ${\bf 3}$. In this case it reduces to the
$SU(2)$ case and the singlets are given in (\ref{ssu5}), (\ref{ssu4}), (\ref{ssu3}) and (\ref{ssu2})

{\it Step 3-2}: If there is an additional singlet in the decomposition (\ref{su2}), then the stability subgroup is $\{1\}$, i.e.~all
spinors are singlets.

{\it Step 3-3}: Additional singlets in the decomposition (\ref{so3}) can either be in the ${\bf 2}$ or in ${\bf 3}$ representations.
If the singlet is in ${\bf 2}$, then the isotropy group is $\{1\}$ and all spinors are invariant.
 If the singlet is in ${\bf 3}$, the isotropy group is $U(1)$
and it reduces to the (\ref{u1}) case above.

The existence of any additional singlets in the decompositions described above have isotropy group $\{1\}$. As a results, all spinors
will be invariant. The results have been  summarized in the table 2.

\subsection{The $Spin(7)\ltimes\bR^9$ series}\label{noncompsingletsapp}

Next let us consider the case of the  $Spin(7)\ltimes\bR^9$ orbit.
 In this case it is best to view the ${\bf 32}$ representation of $Spin(10,1)$ as the sum of the Majorana-Weyl ${\bf 16}^+$ and anti-Majorana-Weyl
${\bf 16}^- $  representations of $Spin(9,1)$.  Under $Spin(7)\subset Spin(9,1)$ the ${\bf 16}^+$ and ${\bf 16}^- $ decompose as
\bea
{\bf 16}^+={\bf 1}\oplus {\bf 7}\oplus {\bf 8}~,~~~{\bf 16}^-={\bf 1}\oplus {\bf 7}\oplus {\bf 8}~.
\eea
Using the null spinor basis, one can easily see that the isotropy group of either singlet is $Spin(7)\ltimes \bR^9$, while
the isotropy group of both is $Spin(7)$. To continue one can proceed as in the heterotic case, which involves the identification
of singlets in ${\bf 16}^+$. Then one also decomposes the ${\bf 16}^- $ under the maximal compact subgroup $K$ of the singlets
of ${\bf 16}^+$. This will give all the singlets of $K$. It turns out that in all cases half of the total number of singlets of $K$ that lie either in
${\bf 16}^+$ or in ${\bf 16}^- $   have
an enhanced isotropy group of the type $K\ltimes\bR^9$. So the isotropy groups of all the spinors in ${\bf 32}$
are of the type $K$ and $K\ltimes\bR^9$, where $K$ is the maximal compact subgroup of the isotropy groups that
appear in the heterotic supergravity. The results are summarized in table 3.

The isotropy groups  $Sp(2)$ (6), $SU(2)^2$ (8), $SU(3)$ (8), $SU(2)$ (10), $U(1)$ (12), $SU(2)$ (16)
can occur as subgroups of  $SU(5)$ and $Spin(7)\ltimes \bR^9$, where $(\cdot)$ denotes $N_H$. However,
$SU(2)\times SU(3)$ appears as a subgroup only of $SU(5)$, while $G_2$ appears
only as a subgroup of  $Spin(7)\ltimes\bR^9$.

It is worth remarking that the lists above include the isotropy groups up to discrete identifications.
There are more possibilities if disconnected
groups are also allowed. For example consider the subgroup $\bZ_2$ generated by $\Gamma_\nat$. Clearly $\Gamma_\nat\in Spin(10,1)$.
The invariant spinors are ${\bf 16}^+$. One can also consider $\bZ_2\times \bR^9$, then the invariant spinors are those in
${\bf 16}^+$ which do not contain the basis form $e_5$. There are many more possibilities.

\appendix{$\Sigma$ groups}
\setcounter{subsection}{0}
  The computation of the $\Sigma$ groups in eleven-dimensional supergravity is similar
 to the one in the heterotic case \cite{hetb}. We shall not give the details of how this Lie algebraic
 computation is done. Instead we shall state explicitly, the generators of the $\Sigma$ groups up to generators of the isotropy
 group $H$. The generators of the $\Sigma$ groups are given in the time-like basis for the $SU(5)$-series and in the null basis
 for the $Spin(7)\ltimes\bR^9$ series \cite{syst}, respectively. For notation and conventions see \cite{ggp, syst}.

\subsection{The $SU(5)$ series}

\subsubsection{$SU(5)$, $N_H=2$}

In this case we find $\Sigma({\cal P})=U(1)$, where the generator of $U(1)$ can be chosen as $i\G^{1\bar 1}$.

\subsubsection{$SU(4)$, $N_H=4$}

Here we get $\Sigma({\cal P})=Spin(2,1)\times U(1)$, where $\mathfrak{spin}(2,1)$ is generated by the real span of
$\G^{05}$, $\G^{0\bar 5}$ and $i\G^{5\bar 5}$,
and the generator of $\mathfrak{u}(1)$ can be chosen as $i\G^{1\bar 1}$.

\subsubsection{$SU(2)\times SU(3)$, $N_H=4$}

For this case we find $\Sigma({\cal P})=SU(2)\times U(1)$, where the generator of $\mathfrak{u}(1)$ can be chosen as $i\G^{3\bar 3}$
and, in the basis
given above, $\mathfrak{su}(2)=\bR<\Gamma^{12}, \Gamma^{\bar1\bar2}, {i\over2}(\Gamma^{1\bar1}+\Gamma^{2\bar2})>$.

\subsubsection{$Sp(2)$, $N_H=6$}

In this case we get $\Sigma({\cal P})=Spin(2,1)\times SU(2)$, where $\mathfrak{spin}(2,1)=\mathfrak{sl}(2,\bR)$ is generated by the real span of $\G^{05}$, $\G^{0\bar 5}$
and $i\G^{5\bar 5}$, and $SU(2)$ acts on ${\cal P}_H$ with the three-dimensional representation. In particular, in the basis
given above   $\mathfrak{su}(2)$ is spanned by
$\Gamma^{12}+\Gamma^{34}, \Gamma^{\bar 1\bar 2}+\Gamma^{\bar 3\bar 4}, {i\over2} (\Gamma^{1\bar1}+\Gamma^{2\bar2}+
\Gamma^{3\bar3}+\Gamma^{4\bar4})$.

\subsubsection{$SU(3)$, $N_H=8$}

Here we get $\Sigma({\cal P})=Spin(4,1)\times U(1)$, where $\mathfrak{spin}(4,1)$ is generated by real span of $\G^{ij}$, with $i,j=0,4,\bar 4,5,\bar 5$,
and the generator of $\mathfrak{u}(1)$ can be chosen as $i\G^{1\bar 1}$.

\subsubsection{$SU(2)^2$, $N_H=8$}

For this case we find $\Sigma({\cal P})=Spin(2,1)\times SU(2)^2$, where $\mathfrak{spin}(2,1)$ is generated by the real span of
 $\G^{05}$, $\G^{0\bar 5}$
and $i\G^{5\bar 5}$, and the two factors of $SU(2)$ act on ${\cal P}_H$ with the three-dimensional representation. In particular,
in the basis given above the two factors of $SU(2)$ are generated by $\mathfrak{su}(2)=\bR<\Gamma^{12}, \Gamma^{\bar1\bar2},
{i\over2}(\Gamma^{1\bar1}+\Gamma^{2\bar2})>$ and $\mathfrak{su}(2)=\bR<\Gamma^{34}, \Gamma^{\bar3\bar4}, {i\over2}(\Gamma^{3\bar3}
+\Gamma^{4\bar4})>$, respectively.

\subsubsection{$SU(2)$, $N_H=10$}

In this case we get $\Sigma({\cal P})=Spin(2,1)\times Sp(2)$, where $\mathfrak{spin}(2,1)$ is generated by the real span of $\G^{05}$,
$\G^{0\bar 5}$ and $i\G^{5\bar 5}$, and $Sp(2)$ acts on ${\cal P}_H$ with the five-dimensional vector representation,
$Sp(2)=Spin(5)$. This can be verified by a direct computation.

\subsubsection{$U(1)$, $N_H=12$}

Here we get $\Sigma({\cal P})=Spin(2,1)\times SU(4)$, where $\mathfrak{spin}(2,1)$ is generated by the real span of $\G^{05}$,
$\G^{0\bar 5}$ and $i\G^{5\bar 5}$, and $SU(4)$ acts on ${\cal P}$ with the real six-dimensional vector representation, $SU(4)=Spin(6)$.
This works exactly as in the corresponding Type I case \cite{hetb} and can easily be seen from previous results by a direct computation.

\subsubsection{$SU(2)$, $N_H=16$}

For this case we find $\Sigma({\cal P})=Spin(6,1)\times SU(2)$, where $\mathfrak{spin}(6,1)$ is generated by the real span of $\G^{ij}$,
with $i,j=0,3,\bar 3,4,\bar 4,5,\bar 5$, and
$SU(2)$ is generated by $\G^{12}+\G^{\bar{1} \bar{2}}$,
$i(\G^{12}-\G^{\bar{1} \bar{2}})$, $i(\G^{1 \bar{1}}+ \Gamma^{2 \bar{2}})$.

\subsection{The $Spin(7)\ltimes\bR^9$ series}

The generators of the $\Sigma$ groups for this series are given in the null basis.

\subsubsection{$Spin(7) \ltimes \bR^9$, $N_H=1$}

For this case, $\Sigma({\cal P}) = Spin(1,1)$, where $\mathfrak{spin}(1,1)$ is generated by   $\Gamma^{+-}$

\subsubsection{$Spin(7)$, $N_H=2$}

For this case, $\Sigma({\cal P}) = Spin(2,1)$, where $\mathfrak{spin}(2,1)$ is generated by
$\Gamma^{+-}, \Gamma^{- \sharp}, \Gamma^{+ \sharp}$

\subsubsection{$G_2 \ltimes \bR^9$, $N_H=2$}

For this case $\Sigma({\cal P}) = Spin(1,1) \times U(1)$, where $\mathfrak{spin}(1,1)$ is
generated by
 $\Gamma^{+-}$ and $\mathfrak{u}(1)$ has generator $(\Gamma^1+\Gamma^{\bar{1}}) \Gamma^\sharp$

\subsubsection{$SU(4) \ltimes \bR^9$, $N_H=2$}

For this case $\Sigma({\cal P}) = Spin(1,1) \times U(1)$, where $\mathfrak{spin}(1,1)$ is generated by
$\Gamma^{+-}$ and $\mathfrak{u}(1)$ is generated by $i (\Gamma^{1 \bar{1}}
+ \Gamma^{2 \bar{2}} + \Gamma^{3 \bar{3}} + \Gamma^{4 \bar{4}})$

\subsubsection{$Sp(2) \ltimes \bR^9$, $N_H=3$}

For this case $\Sigma({\cal P}) = Spin(1,1) \times SU(2)$, where $\mathfrak{spin}(1,1)$ is generated
by $\Gamma^{+-}$ and $\mathfrak{su}(2)$ is generated by
$\Gamma^{12}+ \Gamma^{\bar{1} \bar{2}} + \Gamma^{34} +\Gamma^{\bar{3} \bar{4}},
i(\Gamma^{12}- \Gamma^{\bar{1} \bar{2}} + \Gamma^{34} - \Gamma^{\bar{3} \bar{4}}),
 i (\Gamma^{1 \bar{1}}
+ \Gamma^{2 \bar{2}} + \Gamma^{3 \bar{3}} + \Gamma^{4 \bar{4}})$

\subsubsection{$SU(4)$, $N_H=4$}

For this case $\Sigma({\cal P}) = Spin(2,1) \times U(1)$,
where $\mathfrak{spin}(2,1)$ has generators
$\Gamma^{+-}, \Gamma^{- \sharp}, \Gamma^{+ \sharp}$ and
$\mathfrak{u}(1)$ is generated by
$i (\Gamma^{1 \bar{1}}
+ \Gamma^{2 \bar{2}} + \Gamma^{3 \bar{3}} + \Gamma^{4 \bar{4}})$

\subsubsection{$G_2$, $N_H=4$}

For this case $\Sigma({\cal P}) = Spin(3,1) $, where $\mathfrak{spin}(3,1)$ is generated by
$\Gamma^{+-}, \Gamma^{+ \sharp}, \Gamma^{- \sharp},
\Gamma^+ (\Gamma^1 + \Gamma^{\bar{1}}), \Gamma^- (\Gamma^1 + \Gamma^{\bar{1}}),
\Gamma^\sharp (\Gamma^1 + \Gamma^{\bar{1}})$

\subsubsection{$(SU(2) \times SU(2)) \ltimes \bR^9$, $N_H=4$}

For this case $\Sigma({\cal P}) = Spin(1,1) \times SU(2) \times SU(2) $,
where $\mathfrak{spin}(1,1)$ is generated by $\Gamma^{+-}$ and the two $\mathfrak{su}(2)$ factors are generated by
$\Gamma^{12}+ \Gamma^{\bar{1} \bar{2}}, i (\Gamma^{12}- \Gamma^{\bar{1} \bar{2}}),
i(\Gamma^{1 \bar{1}}+\Gamma^{2 \bar{2}})$ and $\Gamma^{34}+ \Gamma^{\bar{3} \bar{4}}, i (\Gamma^{34}- \Gamma^{\bar{3} \bar{4}}),
i(\Gamma^{3 \bar{3}}+\Gamma^{4 \bar{4}})$, respectively.

\subsubsection{$SU(3) \ltimes \bR^9$, $N_H=4$}

For this case $\Sigma({\cal P}) = Spin(1,1) \times Spin(3) \times U(1) $,
where $\mathfrak{spin}(1,1)$ is generated by $\Gamma^{+-}$, $\mathfrak{spin}(3)$ is generated by
all $\Gamma^{AB}$ for $A,B=1, {\bar{1}}, \sharp$ and
$\mathfrak{u}(1)$ is generated by  $i (
\Gamma^{2 \bar{2}} + \Gamma^{3 \bar{3}} + \Gamma^{4 \bar{4}})$

\subsubsection{$SU(2) \ltimes \bR^9$, $N_H=5$}

For this case $\Sigma({\cal P}) = Spin(1,1) \times Spin(5)$,
where $\mathfrak{spin}(1,1)$ is generated by $\Gamma^{+-}$ and
$\mathfrak{spin}(5)$ is generated by the real span of
$i (\Gamma^{1 \bar{1}}+\Gamma^{2 \bar{2}}),
\Gamma^{12}+\Gamma^{\bar{1} \bar{2}}, i (\Gamma^{12}-\Gamma^{\bar{1} \bar{2}}),
 i (\Gamma^{3 \bar{3}}+\Gamma^{4 \bar{4}}),
\Gamma^{34}+\Gamma^{\bar{3} \bar{4}}, i (\Gamma^{34}-\Gamma^{\bar{3} \bar{4}}),
\Gamma^{13}+\Gamma^{\bar{1} \bar{3}}+\Gamma^{24}+\Gamma^{\bar{2} \bar{4}},
i(\Gamma^{13}-\Gamma^{\bar{1} \bar{3}}+\Gamma^{24}-\Gamma^{\bar{2} \bar{4}}),
\Gamma^{1 \bar{4}}+\Gamma^{\bar{1} 4}+\Gamma^{3 \bar{2}}+\Gamma^{\bar{3} 2},
i(\Gamma^{1 \bar{4}}-\Gamma^{\bar{1} 4}-\Gamma^{3 \bar{2}}+\Gamma^{\bar{3} 2})$.

\subsubsection{$Sp(2)$, $N_H=6$}

For this case $\Sigma({\cal P}) = Spin(2,1) \times SU(2)$,
where $\mathfrak{spin}(2,1)$ is generated by
$\Gamma^{+ \sharp}, \Gamma^{- \sharp}, \Gamma^{+-}$
and $\mathfrak{su}(2)$ is generated by $ \Gamma^{12}
+\Gamma^{\bar{1} \bar{2}}+\Gamma^{34}+\Gamma^{\bar{3} \bar{4}},
i (\Gamma^{12}
-\Gamma^{\bar{1} \bar{2}}+\Gamma^{34}-\Gamma^{\bar{3} \bar{4}}),
i(\Gamma^{1 \bar{1}}+\Gamma^{2 \bar{2}}+\Gamma^{3 \bar{3}}+\Gamma^{4 \bar{4}})$

\subsubsection{$U(1) \ltimes \bR^9$, $N_H=6$}

For this case $\Sigma({\cal P}) = Spin(1,1) \times Spin(6)$,
where $\mathfrak{spin}(1,1)$ is generated by $\Gamma^{+-}$ and $\mathfrak{spin}(6)=\mathfrak{su}(4)$ has
generators
$i (\Gamma^{1 \bar{1}}+\Gamma^{2 \bar{2}}),
\Gamma^{12}+\Gamma^{\bar{1} \bar{2}}, i (\Gamma^{12}-\Gamma^{\bar{1} \bar{2}}),
 i (\Gamma^{3 \bar{3}}+\Gamma^{4 \bar{4}}),
\Gamma^{34}+\Gamma^{\bar{3} \bar{4}}, i (\Gamma^{34}-\Gamma^{\bar{3} \bar{4}}),
i (\Gamma^{1 \bar{1}}+\Gamma^{3 \bar{3}}),
\Gamma^{13}+\Gamma^{\bar{1} \bar{3}}, i (\Gamma^{13}-\Gamma^{\bar{1} \bar{3}}),
\Gamma^{24}+\Gamma^{\bar{2} \bar{4}}, i (\Gamma^{24}-\Gamma^{\bar{2} \bar{4}}),
\Gamma^{2 \bar{3}}-\Gamma^{3 \bar{2}}, i(\Gamma^{2 \bar{3}}+\Gamma^{3 \bar{2}}),
\Gamma^{4 \bar{1}}-\Gamma^{1 \bar{4}}, i(\Gamma^{4 \bar{1}}+\Gamma^{1 \bar{4}})$

\subsubsection{$SU(2) \times SU(2)$, $N_H=8$}

For this case $\Sigma({\cal P}) = Spin(2,1) \times SU(2) \times SU(2)$,
where $\mathfrak{spin}(2,1)$ is generated by
$ \Gamma^{+ \sharp}, \Gamma^{- \sharp}, \Gamma^{+-}$ and the two
$\mathfrak{su}(2)$ factors are generated by $i (\Gamma^{1 \bar{1}}+\Gamma^{2 \bar{2}}),
\Gamma^{12}+\Gamma^{\bar{1} \bar{2}}, i (\Gamma^{12}-\Gamma^{\bar{1} \bar{2}})$
and $i (\Gamma^{3 \bar{3}}+\Gamma^{4 \bar{4}}),
\Gamma^{34}+\Gamma^{\bar{3} \bar{4}}, i (\Gamma^{34}-\Gamma^{\bar{3} \bar{4}})$, respectively.

\subsubsection{$SU(3)$, $N_H=8$}

For this case $\Sigma({\cal P}) = Spin(4,1) \times U(1)$, where $\mathfrak{spin}(4,1)$ is generated by the real span of
$\Gamma^{+-}, \Gamma^{- \sharp}, \Gamma^{+ \sharp},
(\Gamma^{1}+\Gamma^{\bar{1}}) \Gamma^+, i (\Gamma^{1}-\Gamma^{\bar{1}}) \Gamma^+,
(\Gamma^{1}+\Gamma^{\bar{1}}) \Gamma^-, i (\Gamma^{1}-\Gamma^{\bar{1}}) \Gamma^-,
(\Gamma^{1}+\Gamma^{\bar{1}}) \Gamma^\sharp, i (\Gamma^{1}-\Gamma^{\bar{1}}) \Gamma^\sharp,
i \Gamma^{1 \bar{1}}$ and $\mathfrak{u}(1)$ is generated by $i \Gamma^{2 \bar{2}}$.

\subsubsection{$SU(2) \ltimes \bR^9$, $N_H=8$}

For this case $\Sigma({\cal P}) = Spin(1,1) \times Spin(5) \times
SU(2)$, where $\mathfrak{spin}(1,1)$ is generated by $\Gamma^{+-}$,
$\mathfrak{spin}(5)=\mathfrak{sp}(2)$ is generated by
$\Gamma^{12}+\Gamma^{\bar{1}
\bar{2}}, i (\Gamma^{12}-\Gamma^{\bar{1} \bar{2}}), i \Gamma^{1
\bar{1}}, i \Gamma^{2 \bar{2}}, i(\Gamma^{2} - \Gamma^{\bar{2}})
\Gamma^\sharp, (\Gamma^2+\Gamma^{\bar{2}}) \Gamma^{\sharp},
i(\Gamma^{1} - \Gamma^{\bar{1}}) \Gamma^\sharp,
(\Gamma^1+\Gamma^{\bar{1}}) \Gamma^{\sharp}, \Gamma^{1 \bar{2}}-
\Gamma^{2 \bar{1}}, i(\Gamma^{1 \bar{2}}+ \Gamma^{2 \bar{1}})$
and $\mathfrak{su}(2)$ has generators
$\Gamma^{34}+\Gamma^{\bar{3} \bar{4}}, i( \Gamma^{34}-\Gamma^{\bar{3}
\bar{4}}), i(\Gamma^{3 \bar{3}}+\Gamma^{4 \bar{4}})$.

\subsubsection{$SU(2)$, $N_H=10$}

For this case $\Sigma({\cal P}) = Spin(2,1) \times Spin(5)$,
where $\mathfrak{spin}(2,1)$ is generated by
$\Gamma^{+-}, \Gamma^{+ \sharp}, \Gamma^{- \sharp}$,
and $\mathfrak{spin}(5)$ is generated by
$\Gamma^{13}+\Gamma^{\bar{1} \bar{3}}+\Gamma^{24}+\Gamma^{\bar{2} \bar{4}},
i(-\Gamma^{13}+\Gamma^{\bar{1} \bar{3}}-\Gamma^{24}+\Gamma^{\bar{2} \bar{4}}),
\Gamma^{1 \bar{4}}-\Gamma^{4 \bar{1}}-\Gamma^{2 \bar{3}}+\Gamma^{3 \bar{2}},
i(\Gamma^{1 \bar{4}}+\Gamma^{4 \bar{1}}-\Gamma^{2 \bar{3}}-\Gamma^{3 \bar{2}}),
i (\Gamma^{1 \bar{1}}+\Gamma^{2 \bar{2}}),
\Gamma^{12}+\Gamma^{\bar{1} \bar{2}}, i (\Gamma^{12}-\Gamma^{\bar{1} \bar{2}}),
 i (\Gamma^{3 \bar{3}}+\Gamma^{4 \bar{4}}),
\Gamma^{34}+\Gamma^{\bar{3} \bar{4}}, i (\Gamma^{34}-\Gamma^{\bar{3} \bar{4}})$.

\subsubsection{$U(1)$, $N_H=12$}

For this case $\Sigma({\cal P}) = Spin(2,1) \times Spin(6)$,
where $\mathfrak{spin}(2,1)$ is generated by
$\Gamma^{+-}, \Gamma^{-\sharp}, \Gamma^{+ \sharp}$ and
$\mathfrak{spin}(6)$ is generated by $\Gamma^{12}+\Gamma^{\bar{1} \bar{2}},
i(\Gamma^{12}-\Gamma^{\bar{1} \bar{2}}),  \Gamma^{34}+\Gamma^{\bar{3} \bar{4}},
i(\Gamma^{34}-\Gamma^{\bar{3} \bar{4}}),  \Gamma^{24}+\Gamma^{\bar{2} \bar{4}},
i(\Gamma^{24}-\Gamma^{\bar{2} \bar{4}}),  \Gamma^{13}+\Gamma^{\bar{1} \bar{3}},
i(\Gamma^{13}-\Gamma^{\bar{1} \bar{3}}), \Gamma^{1 \bar{4}}-\Gamma^{4 \bar{1}},
i(\Gamma^{1 \bar{4}}+\Gamma^{4 \bar{1}}),  \Gamma^{2 \bar{3}}-\Gamma^{3 \bar{2}},
i(\Gamma^{2 \bar{3}}+\Gamma^{3 \bar{2}}), i(\Gamma^{2 \bar{2}}-\Gamma^{3 \bar{3}}),
i(\Gamma^{1 \bar{1}}-\Gamma^{4 \bar{4}}), i(\Gamma^{1 \bar{1}}+\Gamma^{3 \bar{3}})$.

\subsubsection{$SU(2)$, $N_H=16$}

For this case $\Sigma({\cal P}) = Spin(6,1) \times SU(2)$, where $\mathfrak{su}(2)$ is generated by
$i (\Gamma^{3 \bar{3}}+\Gamma^{4 \bar{4}}),
\Gamma^{34}+\Gamma^{\bar{3} \bar{4}}, i (\Gamma^{34}-\Gamma^{\bar{3} \bar{4}})$, and
$\mathfrak{spin}(6,1)$ is generated by the real span of $\Gamma^{AB}$ for $A,B=+, -, \sharp, 1, \bar{1}, 2, \bar{2}$.

\subsubsection{$\bR^9$, $N_H=16$}

For this case $\Sigma({\cal P}) = Spin(1,1) \times Spin(9)$, where
$\mathfrak{spin}(1,1)$ is generated by $\Gamma^{+-}$ and $\mathfrak{spin}(9)$ is generated
by the real span of $\Gamma^{AB}$ for $A,B=1,2,3,4,{\bar{1}}, {\bar{2}},
{\bar{3}}, {\bar{4}}, \sharp$.

\end{document}